\documentclass[11pt,a4paper]{article}
\def\bea{\begin{eqnarray}}
\def\eea{\end{eqnarray}}
\def\nn{\nonumber}
\def\beq{\begin{equation}}
\def\eeq{\end{equation}}
\def\ba{\beq\new\begin{array}{c}}
\def\ea{\end{array}\eeq}
\def\be{\ba}
\def\ee{\ea}

\makeatletter
\newdimen\normalarrayskip              
\newdimen\minarrayskip                 
\normalarrayskip\baselineskip
\minarrayskip\jot
\newif\ifold             \oldtrue            \def\new{\oldfalse}
\def\arraymode{\ifold\relax\else\displaystyle\fi} 
\def\eqnumphantom{\phantom{(\theequation)}}     
\def\@arrayskip{\ifold\baselineskip\z@\lineskip\z@
     \else
     \baselineskip\minarrayskip\lineskip2\minarrayskip\fi}
\def\@arrayclassz{\ifcase \@lastchclass \@acolampacol \or
\@ampacol \or \or \or \@addamp \or
   \@acolampacol \or \@firstampfalse \@acol \fi
\edef\@preamble{\@preamble
  \ifcase \@chnum
     \hfil$\relax\arraymode\@sharp$\hfil
     \or $\relax\arraymode\@sharp$\hfil
     \or \hfil$\relax\arraymode\@sharp$\fi}}
\def\@array[#1]#2{\setbox\@arstrutbox=\hbox{\vrule
     height\arraystretch \ht\strutbox
     depth\arraystretch \dp\strutbox
     width\z@}\@mkpream{#2}\edef\@preamble{\halign
\noexpand\@halignto
\bgroup \tabskip\z@ \@arstrut \@preamble \tabskip\z@ \cr}%
\let\@startpbox\@@startpbox \let\@endpbox\@@endpbox
  \if #1t\vtop \else \if#1b\vbox \else \vcenter \fi\fi
  \bgroup \let\par\relax
  \let\@sharp##\let\protect\relax
  \@arrayskip\@preamble}
\def\eqnarray{\stepcounter{equation}%
              \let\@currentlabel=\theequation
              \global\@eqnswtrue
              \global\@eqcnt\z@
              \tabskip\@centering
              \let\\=\@eqncr
              $$%
 \halign to \displaywidth\bgroup
    \eqnumphantom\@eqnsel\hskip\@centering
    $\displaystyle \tabskip\z@ {##}$%
    \global\@eqcnt\@ne \hskip 2\arraycolsep
         $\displaystyle\arraymode{##}$\hfil
    \global\@eqcnt\tw@ \hskip 2\arraycolsep
         $\displaystyle\tabskip\z@{##}$\hfil
         \tabskip\@centering
    &{##}\tabskip\z@\cr}
\begingroup\ifx\undefined\newsymbol \else\def\input#1 {\endgroup}\fi
\input amssym.def \relax
\input amssym
\newfont{\hr}{msbm10}
\newfont{\ams}{msam10}

\font\numbers=cmss12
\font\upright=cmu10 scaled\magstep1
\def\stroke{\vrule height8pt width0.4pt depth-0.1pt}
\def\topfleck{\vrule height8pt width0.5pt depth-5.9pt}
\def\botfleck{\vrule height2pt width0.5pt depth0.1pt}
\def\Zmath{\vcenter{\hbox{\numbers\rlap{\rlap{Z}\kern 0.8pt\topfleck}\kern 2.2pt
                   \rlap Z\kern 6pt\botfleck\kern 1pt}}}
\def\Qmath{\vcenter{\hbox{\upright\rlap{\rlap{Q}\kern
                   3.8pt\stroke}\phantom{Q}}}}
\def\Nmath{\vcenter{\hbox{\upright\rlap{I}\kern 1.7pt N}}}
\def\Cmath{\vcenter{\hbox{\upright\rlap{\rlap{C}\kern
                   3.8pt\stroke}\phantom{C}}}}
\def\Rmath{\vcenter{\hbox{\upright\rlap{I}\kern 1.7pt R}}}
\def\Z{\ifmmode\Zmath\else$\Zmath$\fi}
\def\Q{\ifmmode\Qmath\else$\Qmath$\fi}
\def\N{\ifmmode\Nmath\else$\Nmath$\fi}
\def\C{\ifmmode\Cmath\else$\Cmath$\fi}
\def\R{\ifmmode\Rmath\else$\Rmath$\fi}

\newcounter{app}

\def\app{\setcounter{equation}{0}
\def\theequation{\Alph{app}.\arabic{equation}}\par
   \addvspace{4ex}
   \@afterindentfalse
  \secdef\@app\@dapp}

\newcommand\@app{\@startsection {app}{1}{0ex}%
                                   {-3.5ex \@plus -1ex \@minus -.2ex}%
                                   {2.3ex \@plus.2ex}%
                                   {\normalfont\Large\bf}}
\def\@dapp#1{%
{\parindent \z@ \raggedright  \bf #1}\par\nobreak}
\def\l@app#1#2{\ifnum \c@tocdepth >\z@
    \addpenalty\@secpenalty
    \addvspace{1.0em \@plus\p@}%
    \setlength\@tempdima{8em}%
    \begingroup
      \parindent \z@ \rightskip \@pnumwidth
      \parfillskip -\@pnumwidth
      \leavevmode \bfseries
      \advance\leftskip\@tempdima
      \hskip -\leftskip
      #1\nobreak\hfil \nobreak\hb@xt@\@pnumwidth{\hss #2}\par
    \endgroup\fi}
\newcounter{sapp}[app]

\def\sapp{\def\theequation{\Alph{app}.\arabic{equation}}
\par
\@afterindentfalse
  \secdef\@sapp\@dsapp}
\newcommand{\@sapp}{\@startsection{sapp}{2}{\z@}%
                                     {-3.25ex\@plus -1ex \@minus 
-.2ex}%
                                     {1.5ex \@plus .2ex}%
                                     {\normalfont\large\bfseries}}

\def\@dsapp#1{%
{\parindent \z@ \raggedright  \bf #1
}\par\nobreak}
\newcommand{\l@sapp}{\@dottedtocline{2}{1.5em}{2.3em}}

\def\2{{1\over 2}}
\def\N2{${\cal N}=2$}

\def\be{ \begin{eqnarray} }
\def\ee{ \end{eqnarray} }

\def\bea{\begin{eqnarray}}
\def\eea{\end{eqnarray}}
\def\nn{\nonumber}

\def\beq{\begin{equation}}
\def\eeq{\end{equation}}
\def\ba{\beq\new\begin{array}{c}}
\def\ea{\end{array}\eeq}
\def\be{\ba}
\def\ee{\ea}

\title{
The Dijkgraaf-Vafa prepotential
\\
in the context of general Seiberg-Witten theory
}

\author{
H.Itoyama$^{1}$ and A.Morozov$^{2}$ \
\\ \normalsize \em $^{1}$ 
Department of Mathematics and Physics, Osaka City University, Osaka, Japan
\\
\normalsize \em $^{2}$
Institute of Theoretical and Experimental Physics, Moscow}

\date{November, 2002}

\begin{document}

\maketitle

\vspace{-8.2cm}

\begin{center}
\hfill OCU-PHYS 194\\
\hfill ITEP/TH-57/02\\
\hfill hep-th/0211245
\end{center}

\vspace{5.5cm}

\begin{abstract}
We consider the prepotential of Dijkgraaf and Vafa (DV) as one
more (and in fact, singular) example of the Seiberg-Witten (SW)
prepotentials and discuss its properties from this perspective.
Most attention is devoted to the issue of complete system
of moduli, which should include not only the sizes of the
cuts (in matrix model interpretation), but also their
positions, i.e. the number of moduli should be almost
doubled, as compared to the DV consideration.
We introduce the notion of regularized DV system
(not necessarilly related to matrix model) and discuss 
the WDVV equations.
These definitely hold before regularization is lifted,
but an adequate limiting procedure, preserving all ingredients
of the SW theory, remains to be found.
\end{abstract}


\section{Introduction}

The recent papers \cite{DV} of R.Dijkgraaf and C.Vafa (DV)
(see \cite{preh,CIV} for the prehistory and
the follow-up in \cite{ChM}-\cite{followup}) have
attracted a new attention to the {\it unfinished} subject of
the Seiberg-Witten (SW) theory \cite{SW1}-\cite{SU}, 
its relation \cite{GKMMM} to integrability \cite{tauf}-\cite{qutauf} 
in general and to that of matrix models \cite{UFN3} in particular.
The whole story is about the old claim \cite{UFN2}
that "effective actions" (functions of coupling constants and
background fields), obtained after functional integration
over all fields, are the $\tau$-functions \cite{tauf,qutauf} of classical
(if fields are integrated out completely) or quantum
(if some background fields are preserved and considered as
subjects of further averaging, i.e. as operators)
integrable hierarchies. 
The SW theory deals with specific situation, a boundary between
classical and quantum, 
when the non-integrated fields are essentially the vacuum values,
parametrizing non-trivial moduli spaces of vacua in SUSY theories.
As usual, the classical moduli spaces 
are essentially {\it quasiclassical} objects
(in particular, possess distinguished symplectic structures), 
and the corresponding
effective actions (written in terms of {\it prepotentials})
should possess interpretation as "quasiclassical $\tau$-functions".
Empirically this is indeed the case (numerous examples are worked
out since the claim was first put forward in the
context of SW theory in \cite{GKMMM}, see 
\cite{intfollowup}-\cite{intfollowup2}),
but the general theory is still lacking, largely because of the
lack of a relevant self-contained definition of a
{\it quasiclassical $\tau$-function} 
(in varience with the ordinary $\tau$-function, which can be described
in group-theory terms and/or as satisfying peculiar bilinear 
Hirota equations. See a discussion in \cite{genHir}).
All known definitions, including the hierarchy of Whitham equations
\cite{Whith,IM2,GMMM}, 
refer to additional data, like solution of original integrable
equations and original $\tau$-function, of which the prepotential is
a "quasiclassical" approximation, 
or like the SW/Hitchin "fibrations" of complex manifolds over moduli
spaces.
Such definitions, though very important, especially for applications, 
are not fully satisfactory, because one and the same "quasiclassical"
(in fact, the {\it low-energy}, in most applications) 
limit arises for different original systems
(moreover, original system can have different quasiclassical limits
(phases)), 
while the SW/Hitchin pattern should finally arise as a {\it solution}
to some problem, rather than serve as a definition of it.

So far, the two most promising approaches to an {\it a priori}
definition of prepotentials (quasiclassical $\tau$-function)
-- as a peculiar class of special functions --
are to consider them as solutions of some universal (like Hirota)
equation {\it or} to represent them as series or integrals of some 
special type.

The so far best achievement on the first route was 
the formulation of (generalized) "spherical" WDVV equations \cite{WDVV} 
in \cite{MMM1} (see the followup in \cite{MMM2}-\cite{KMWZ}).
Unfortunately, there is more and more evidence that in their
original ("spherical") form, suggested in \cite{MMM1}, these
equations are true only for the {\it hyperelliptic} prepotentials,
while their generalization, adequate to cover at least the elliptic
level, remain unknown.\footnote{
Recently, the old example of Calogero system (associated with the
$N=2$ SUSY deformation of the $N=4$ SUSY YM), where the breakdown
of the {\it spherical} WDVV eqs. of \cite{MMM1} was first observed
\cite{MMM3}, got supplemented by another elliptic prepotential \cite{DHKS}
(the Leigh-Strassler deformation \cite{LS} of the same $N=4$ SUSY YM),
which, according to the proof in \cite{BMMM}, can {\it not} 
satisfy the {\it spherical} WDVV. (Ref.\cite{BMMM} lists all the
solutions of WDVV equations of the form $f(a_i - a_j)$, and there
is no eliptic solution, like 
$\zeta(a_i - a_j + b) + \zeta(a_i - a_j - b)$,
announced to be a prepotential of LS deformation in ref.\cite{DHKS}.)
}

Taking the second route, one naturally arrives at considering
matrix models, which: are, at large $N$, the natural approximations to
ordinary QFT models, like Yang-Mills theories or models, 
based on Kac-Moody algebras 
(like conformal models of the WZNW family \cite{WZNW});  
provide an {\it integrable}
approximation (in the sense that their partition functions are 
$\tau$-functions at all $N$ \cite{MaMo,UFN3});
and since \cite{quclaMaMo} are known to give rise to
quasiclassical $\tau$-functions.
The DV theory \cite{DV} is a new important step in this
direction.

In ref.\cite{DV}, in addition to important results for other
branches of field theory,
a {\it new} example of the SW theory is suggested, with

\be
dS_{DV} = \sqrt{P_n^2(x) + f_{n-1}(x)}dx
\ee
(motivated by the studies of SW theory in relation to Calabi-Yau
manifolds \cite{preh} -- the most natural arena for SW theory to act, 
 (e.g.\cite{IM2} ),
for which the prepotential can be found {\it explicitly}
in the form of a double integral.
THis example appears to coincide with the spherical large-$N$
 limit of partition function of Hermitean
1-matrix model.
This fact at last provides a long-awaited bridge between
matrix models and SW theory.

However, the bridge is still somewhat fragile, despite being
heavily explored, exploited and strengthened by more and more examples.
In our opinion, there are  questions
of principal importance of the following kind which
 have escaped  these studies so far.

First, it is still unclear what happens to the correspondence
for other examples of SW theory, where prepotentials are not
(yet?) representable as double integrals. In fact, this subject
{\it is} getting attention (see, e.g.\cite{DGKV}) and, hopefully,
will be clarified in the near future.

Second, the SW prepotential acquires its full meaning only when
its peculiar arguments, ( which we refer to as {\it flat} moduli
  in this paper) are carefully specified.
Unfortunately, no clear definition (neither conceptual nor
technical) of {\it all} flat moduli is yet known in  matrix models.
(Below in this paper we shall see that there are actually two
types of moduli in the DV theory, $S_i$ and $T_i$. Of these,
$S_i$ has an interpretation as the number of eigenvalues, concentrated
on the $i$-th cut, while the matrix-model definition of $T_i$ is 
still lacking.)  
In fact, the very theory of continuum limit of matrix models
from the perspective of integrable systems is not well-developed.
Since \cite{MaMo,quclaMaMo} the subject has not attracted much
attention.

Third, no {\it a priori} characterization such as a
new universal equation for the prepotential is yet
provided. We shall see that the original DV prepotential
(as a function of {\it all} moduli, $S_i$ and $T_i$) has some chances 
to satisfy just the usual {\it spherical} WDVV equations of
\cite{MMM1} (with the usual problem expected for elliptic
prepotentials, like the one in \cite{DHKS}). Moreover, even
in this hyperelliptic case the {\it raison d'etre} for the
WDVV equations from the matrix-model perspective remains
obscure (from the SW theory perspective the reason is the
existence of peculiar closed algebra of 1-forms \cite{MMM2}).

In the present paper we suggest to start the thorough study
of DV prepotentials {\it per se}, 
irrespective of their exciting applications
to Yang-Mills and string models,
and make some first steps in this direction.
Namely, after a brief discussion of 
generic SW theory and of the DV suggestion in s.2 and 3 respectively,
we introduce in s.4 a {\it regularized} version of DV theory,
-- actually, the most direct generalization of original
ansatz of ref.\cite{SW1}, --
which does not have direct matrix-model representation, but
instead deals with smooth hyperelliptic curves and 
{\it holomorphic} differentials on them.
We define the full set of {\it flat} moduli, the prepotential
(in implicit form, as usual in the SW theory) and
residue formula \cite{resfor} for its third derivatives.
According to the general arguments of \cite{MMM2,MMM3} this
(hyperellipticity, holomorphicity and residue formula) 
is enough to prove the validity of WDVV equations.
In s.5 we briefly discuss the transition to unregularized
DV system and the calculation of the CIV-DV prepotential.
Much remains to be done, however, in order to come to
appropriate understanding of the subject.

\section{General Seiberg-Witten (SW) theory}

Generically, the SW theory \cite{SW1,SW23} 
(in complex dimension one)
includes the following ingredients:

Input data consist of a special family of 
{\it spectral} Riemann surfaces
= complex curves
(with or without singularities) and (a homotopical class of)
a meromorphic 1-differential $dS$ on every surface  with the
property, that its moduli derivatives are {\it less singular}
than $dS$ itself. In practice all the relevant examples can be
considered as the limiting cases of the {\it regularized} ones,
where the requirement is just that the moduli-derivatives of
$dS$ are {\it holomorphic} 1-differentials 
(with no poles at all)\footnote{
Note that $dS$ itself need not be holomorphic even in the 
regularized setting.
}.

Given such a SW family one can further make the following steps.

-- Introduce specific {\it flat} coordinates on the moduli space
(of the SW curves) by integrating $dS$ along the 
$A_p$ contours, $p$ runs from $1$ to $g$, the
genus of the curve (which is the same for the entire family),

\be
a_p = \oint_{A_p} dS;
\label{flatmoddef}
\ee

-- Introduce specific (non-single-valued) function on the moduli
space -- the {\it prepotential} ${\cal F}(a)$ by identifying its 
first $a_p$-derivatives with the integrals of $dS$ along the
conjugated $B_p$ contours,

\be
\frac{\partial {\cal F}(a)}{\partial a_p} = b_p = \oint_{B_p} dS;
\label{foveradef}
\ee 
The self-consistency of such definition, i.e. the symmetricity
of the matrix

\be
T_{pq} = 
\frac{\partial b_p}{\partial a_q} = T_{qp},
\ee 
is guaranteed by the requirements, imposed on the moduli dependence
of $dS$. Actually, $T_{pq}$ is the (always symmetric)
{\it period matrix} of the curve, and derivatives of $dS$ over
{\it flat} moduli, 

\be
d\omega_p = \frac{\partial dS}{\partial a_p}
\ee  
are {\it canonical} 
1-differentials for the  set of $A$-contours, selected in
(\ref{flatmoddef}), i.e.
$\oint_{A_p} d\omega_q = \delta_{pq}$. Moreover, as already
stated, the SW family can be always {\it regularized} in such
a way that $d\omega_p$ are {\it holomorphic} canonical 1-differentials.

-- The SW fibration of complex curves over the SW moduli space
possesses a distinguished symplectic structure

\be
\delta dS = \sum_p \delta a_p\wedge d\omega_p.
\label{syms}
\ee
It reflects the connection between SW theory and
integrable systems \cite{GKMMM},
\cite{intfollowup}-\cite{intfollowup2} (actually, $dS$ is an 
eigenvalue of the Lax 1-form).

-- The third derivatives of the prepotential with respect to the {\it flat}
moduli, which are the first moduli-derivatives of the period
matrices, are given by residue formulas \cite{resfor,MMM1}, like

\be
F_{pqr} = \frac{\partial^3 {\cal F}(a)}
{\partial a_p \partial a_q \partial a_r} = 
\frac{\partial T_{pq}}{\partial a_r}
\sum_{dS=0} res\ \frac{d\omega_p d\omega_q d\omega_r}{ddS},
\label{resfordef}
\ee
where quadratic differential $ddS$ is 
intimately related to the symplectic structure $\delta dS$,
and sum goes over residues at all the zeroes of $dS$.

-- If a meaning can be given to a closed associative algebra of holomorphic 
forms modulo $ddS/dS$ and residue formula is rewritten in terms of a sum
over {\it its} zeroes (this proves possible at least for all 
known examples of {\it hyperelliptic} SW families), eq.(\ref{resfordef})
can be used to derive the WDVV equations 
\cite{WDVV}-\cite{KMWZ}
for the prepotential ${\cal F}(a)$,
 
\be
\check{\cal F}_p \check{\cal F}_q^{-1} \check{\cal F}_r = 
\check{\cal F}_r \check{\cal F}_q^{-1} \check{\cal F}_p
\label{WDVVeqs}
\ee 
for all triples $p,q,r$. Here $\check{\cal F}_p$ denotes a 
(symmetric) matrix with the entries 

\be
(\check{\cal F}_p)_{qr} = F_{pqr} = \frac{\partial^3 {\cal F}(a)}
{\partial a_p \partial a_q \partial a_r}
\ee 
Since matrices $\check{\cal F}$ are symmetric, eqs.(\ref{WDVVeqs}) are 
equivalent to the symmetricity of the matrix 
$ \check{\cal F}_p \check{\cal F}_q^{-1} \check{\cal F}_r$.
In fact, it is enough to check the equations for a given $q$
(and {\it all} $p,r$), then they automatically hold for all values of $q$.
If the set of moduli of the SW system is sufficiently large for
the WDVV eqs. to hold, we call the system {\it complete}.

-- The prepotential can be further extended to a larger moduli
space. In the simplest case of such extention \cite{IM2,GMMM} additional 
moduli are located at a single point $x_0$ on the spectral curve,
where $dS$ developes a singularity (generically,
an essential\footnote{
A natural interpretation/generalization of $dS$ is in terms
of correlators of free fields on a Riemann surface (either
on the spectral surface itself or, better, on the {\it bare}
spectral surface \cite{IM1} of which the former one is
a ramified covering). Introduction of essential singularities
is then inspired by the theory of Baker-Akhiezer functions,
 which explains -- and further inspires -- many
of links to integrable systems. Despite its importance for
further developements of the subject (especially, for connection
to KP/Toda-like $\tau$-functions\cite{tauf}, 
to matrix models at finite $N$
\cite{UFN3} and to considerations of ref.\cite{LMNS}-\cite{F}),
this subject lies beyond the scope of the present paper, which is
devoted entirely to quasiclassical $\tau$-functions (prepotentials).
} one), 
and (\ref{flatmoddef}), (\ref{foveradef}) get supplemented by

\be
\tilde T_k = res_{x_0} (x-x_0)^kdS(x), \nn \\
\frac{\partial {\cal F}(a)}{\partial \tilde T_k} = 
\frac{1}{k}res_{x_0} (x-x_0)^{-k}dS(x)
\label{Whmoduli}
\ee 
with all $k\geq 0$
(here some particular choice of coordinates in the vicinity
of $x_0$ is implied, also logarithm is implicit in the
formula for the $T_0$ derivative). 
If the singularity at $x=x_0$ is
not an essential one, but just a finite-order pole, the number
of additional moduli is finite.
At least in this case the singularity can be considered as
a result of degeneration of the non-singular moduli like
(\ref{flatmoddef}) -- this brings us back to the notion of 
{\it regularized} SW system.\footnote{
An important example is description of fundamental matter
in SW theory \cite{Masf}. When some $2N_f$ branching points of the
hyperelliptic spectral curve
$$
y^2 = \tilde P_{\tilde N_c}^2(x) - \Lambda^{2\tilde N_c}
$$
collide pairwise (this happens at the 
hypersurfaces of vanishing monopole masses  
in the original moduli space), 
$$
\tilde P_{\tilde N_c}^2(x) - \Lambda^{2\tilde N_c} =
Q_{N_f}^2(x)F_{2N_c}(x), \ \ 
Q_{N_f}(x) = \prod_{\nu =1}^{N_f}(x-m_\nu),
$$
the newly emerging polynomial of degree $2N_c = 2\tilde N_c -
2N_f$ can be represented as
$$
F_{2N_c}(x) = P_{N_c}^2(x) - \Lambda^{2(N_c-N_f)}Q_{N_f}(x)
$$
with some new polynomial $P_{N_c}(x)$ of degree $N_c$.
The corresponding  $dS = \lambda \frac{dw}{w} = x d\log(P+y)$ 
\cite{Masf} has simple poles at the points $m_\nu$, i.e.
describes a singular SW model.
Original "pure gauge" SW model with $\tilde N_c = N_c + N_f$
can be considered as a {\it regularization} of the singular SW model
with $N_c$ colors and $N_f$ flavors, obtained by the
blowing up of degenerated handles. This also explains, why
the pole positions ("masses") $m_\nu$ should necessarily
be included into the set of SW moduli \cite{MMM3}.
}
Instead of taking simple singularities, one can consider 
{\it larger} sets of singular points $x_0$ and, further, make
them form {\it continuous} sets, like cuts, closed contours
and entire areas inside the contours.
Certain developement in this direction is due to the series of
papers \cite{KMWZ}, in particular, the WDVV eqs. are shown
to survive this type of generalizations.
The DV theory \cite{DV} actually analyzes the case of
the multicomponent contours (multicuts).

-- On infinitely extended moduli space the prepotential still
satisfies integrable equations of {\it Whitham}
 {\it hierarchies} \cite{Whith}, and finally can be considered as a
peculiar {\it quasiclassical} limit of the logarithm of a 
$\tau$-{\it function}. With no surprise, this procedure is
believed to be in parallel with the occurence of
the quasiclassical $\tau$-functions in the {\it planar}
large-$N$ limit of the matrix models (which, before taking the
limit, are ordinary KP-like $\tau$-functions
\cite{UFN3}). 
This parallel was recently pushed much closer to concrete
statements in \cite{DV} and, even further, in \cite{ChM}.
What is most interesting in this developement -- from the point of
view of prepotential theory -- is discovery of explicit expressions
(in the form of a double integral, actually, the quasiclassical
partition of a matrix model) for the prepotentials of certain
SW systems.

\section{DV Theory}

\subsection{Generalities}

The DV theory attracts a lot of attention because of its non-trivial
message to physics of Yang-Mills theories: 
the two seemingly different procedures,-- 
(i) the non-perturbative analysis of the low-energy actions, performed
(at least implicitly) with the help of sophisticated instanton calculus, 
and
(ii) the rather naive analysis of the zero-mode sector (reduction to
$d=0$) in the large-$N$ limit, giving rise to the well-known
planar matrix models \cite{MaMop},--
provide the same quantities. While being not very surprising
conceptually (especially to those who believe in universality
classes of effective actions and their integrable properties
which reflect this universality and hidden symmetries),
this is a very concrete statement, opening the field for detailed
analysis and raising a lot of challenging questions.

Our goal is somewhat orthogonal to this line of thought
about DV theory. We rather concentrate on its value for purely theoretical
purposes, namely for the study of quasiclassical $\tau$-functions
and their properties.
The main achievement of the DV theory from the point of view of
generic SW theory is that it seems to provide a set of examples 
-- matrix models in the planar continuous limit --
where all of the four interrelated ingredients of the SW theory
(the set of curves, the form $dS$, the flat moduli $a_i=\oint_{A_i}dS$
and the prepotential ${\cal F}(a_i)$) have direct interpretation.
In all previous examples of SW theory only some of these ingredients
appeared in a natural way, while the others were built with the
help of the SW theory itself:
in SUSY YM theories at low energies the natural things are
moduli and prepotential, while spectral curves are hidden structures
(revealed in \cite{SW1});
in brane patterns the natural things are instead the Lax forms 
(describing the shape of interacting branes, \cite{Wbr,MartMM})
and their spectral curves;
in Calabi-Yau studies again the curves are natural, as describing
the geometry of Calabi-Yau 3-folds near the singularities,
$dS$ then arises as restriction of the distinguished 3-form, 
but the prepotential should be introduced as some extra quantity,
characterizing an associated SUSY sigma-model (in fact, one can
consider this association, described sometime in terms of the
"special geometry" \cite{speg} as another, not too different, formulation
of SW theory); etc.
The planar continuum limit of matrix models, however, seems to
provide everything simultaneously:
$dS$ is associated with the density $\rho(\lambda)d\lambda$
of the eigenvalues, the curves appear from {\it algebraic}
equations of motion for $\rho(\lambda)$, the flat moduli
are just the fractions of eigenvalues on every cut and
the prepotential is nothing but the free energy $\log Z_{planar}$
of the model. 
The achievement of simultaneous description of all the ingredients
bears an immediate fruit: in
so obtained SW models there is a straightforward
double-integral representation for the
prepotential

\be
{\cal F}(a_i) = \sum_{i,j} \oint_{A_i}\oint_{A_j}
dS(x) dS(x') G(x,x'), \ \
a_i = \oint_{A_i}dS
\label{prepdi}
\ee
with certain kernel (Green's function) $G(x,x')$.

Unfortunately, there are still clouds over this idyllic picture. 
The most important one for the purposes of the present paper
is somewhat unnatural restriction on the set of moduli. Obviously,
the actual moduli of the planar-limit
matrix model include more than just the numbers of eigenvalues
on cuts; there are also cut's positions. (The lengths of the cuts
 are expressible through the numbers of eigenvalues, but 
the positions of their centers remain independent variables).
Without inclusion of all these moduli, there is no grounded hope
of obtaining a {\it complete} SW system, in particular of obtaining
any universal equations, like WDVV, for the prepotential.
We consider our attempts to introduce back these "lost" moduli in
the remaining sections below.
It looks like, after being introduced, they can indeed restore the
WDVV equations.
This is absolutely obvious for a "regularized DV" system, to be
introduced and analyzed in s.4 below,
but instead the simple relation to matrix models can be lost after
regularization.
From this point of view it becomes clear that the original DV
systems, coming from matrix models, are in fact examples of
{\it singular} SW models, where many of moduli get localized
at particular points (and then easily overlooked, as it happens
in most analyses of the situation in the literature).
Careful analysis of the singular limit is rather difficult,
even with the shortcuts, making use of the prepotential theory
\cite{IM2,GMMM}, see s.5 below.

Before proceeding to describe these results in the
next sections 4 and 5, we briefly summarize the statements suggested
in \cite{DV} with some emphasis on the questions, which,
in our opinion, deserve further investigation.

\subsection{From matrix model to a SW model}

The recipe is:

1) Take some matrix model,

\be
Z_N = \int  e^{-Tr\ {\cal L}(\Phi)}d\Phi
\label{ormm}
\ee
(There can be many matrix-valued $\Phi$-variables and multitrace
operators can appear in the action).

2) Rewrite it in terms of eigenvalues:

\be
Z_N =  \int e^{-{\cal S}\{\lambda\}} \prod_s  d\lambda_s
\ee
(For non-eigenvalue model, the integral over unitary matrices
can not be taken in any final form. Then ${\cal S}\{\lambda\}$
is a formal series, derivable term by term with the help of
unitary integrals, see \cite{MSha} for some relevant formulas).

3) Rewrite it further in terms of an integral over eigenvalue densities 
$\rho(\lambda)d\lambda = \sum_s \delta(\lambda - \lambda_s)d\lambda$.

\be
Z_N = \int  e^{-S\{\rho(\lambda)\}} D\rho(\lambda)
\ee
Note, that the density $\rho(\lambda)d\lambda$ is actually a 1-form.
This step can be considered unnecessary for the rest
of the DV construction,  although it may be still important, at least
conceptually.

4) Write down a classical equation of motion for $\rho(\lambda)$
(in principle, it can be extracted directly from
${\cal S}\{\lambda\}$, introduced at step 2),

\be
Diff\{\rho(\lambda)\} = 0
\label{eqnrho}
\ee
Generically, this is a differential equation (a loop equation
or the Ward identity, of primary importance in the theory of
matrix models \cite{MaMop,loop,MaMo,UFN3}).

5) In fact, there are two different forms of eq.(\ref{eqnrho}),
related by differentiation. DV prescription is to choose
the option with teh derivative taken and {\it ignore} the constraints, imposed
on integration constants (which would require
that the level of Fermi surface is the same for all
cuts). This step is crucial for the DV
construction, and this is where it deviates from the usual
studies of planar matrix models, allowing more solutions
(more moduli); see \cite{ChM} for more details.

6) In a straightforward large-$N$  limit, equation 
(\ref{eqnrho}) becomes
{\it algebraic} and its solution, $\rho_0(\lambda)d\lambda$,
acquires interpretation in terms of a Riemann surface
(complex spectral curve). The moduli of this solution, --
positions of ramification points,-- become
parameters, e.g. the positions of the
cuts.
At this step the information about multitrace terms in
original action in (\ref{ormm}) can be lost -- as usual in the
naive large-$N$ limit.

7) The classical action, evaluated at this solution,

\be
{\cal F}(moduli) = S\{\rho_0(\lambda)d\lambda\}
\ee
when considered as a function of moduli, should be
a prepotential of some SW model.

8) The commonly adopted postulates are:

-- The SW structure is defined by 
the 1-form $\rho_0(\lambda)d\lambda$:
\be
dS_{DV} = \rho_0(\lambda)d\lambda
\ee 

-- Moduli are just the integrals of $\rho_0(\lambda)d\lambda$
along the cuts.

\bigskip

There are some immediate obvious  questions to be asked about these statements.

What are the moduli? Why only "lengths" of the cuts, but not their
"positions" are included in the DV considerations? 
In other words, what prevents one from considering all ramification points
as moduli of the solutions to the algebraic limit of (\ref{eqnrho})?

Why does $\rho_0(\lambda)d\lambda$ provide a SW structure? 
Why are moduli derivatives 
 holomorphic (in fact, not quite!) on the spectral curve?
Why 
\be
\frac{\partial S\{\rho_0(\lambda)d\lambda\}}{\partial (moduli)} =
B-periods\ of\ \rho_0(\lambda)d\lambda?
\ee
In fact, in the examples, the arising SW structure is {\it singular} in
the  sense that
certain {\it mero}morphic differentials should be included, and
some entries of the period matrix diverge. This requires some kind
of {\it regularization}. However, the naive cut-off procedure,
included into the DV recipe, while providing a fast short-cut to
physical applications, can (and does) cause problems in theoretical
investigation of the relevant structures and hidden symmetries.
(This group of questions is partly addressed in ref.\cite{ChM}.)
Is there any independent self-contained characterization of
the prepotential $S\{\rho_0(\lambda)d\lambda\}$ (like a WDVV-like equation)?

\subsection{From SW model to a matrix model}

Not too much seems to be currently known on 
the subject of this subsection.
Leaving aside many obvious, but more detailed questions,
at the beginning it is enough to ask:

Why should the
prepotential be expressed by eq.(\ref{prepdi})
and what is the relevant kernel (Green function)
$G(x,x')$?

In general SW theory, whenever a prepotential is
homogeneous of degree $2$, 

\be
\sum_I a_I \frac{\partial{\cal F}}{\partial a_I} = 2{\cal F},
\ee
one always has  \cite{IM2}:

\be
{\cal F}(a_I) = \frac{1}{2}\sum_{I} \oint_{A_I}dS \oint_{B_I}dS,
\ \ a_I = \oint_{A_I}dS
\label{prepgen}
\ee
At the first glance this can seem to be the double-integral
representation of interest. As a matter of principle, it is;
however, there are two essential differences between eqs.
(\ref{prepdi}) and (\ref{prepgen}), which allow to consider
(\ref{prepdi}) as a significant step forward as compared to 
(\ref{prepgen}) -- if it can be extended to sufficiently 
general circumstances.

First, in most examples, ${\cal F}(a_I)$ is not really
homogeneous of order $2$, because of {\it anomalies},
responsible for logarithmic contributions to the prepotential.
(In matrix models it is common to attribute these to the
"volume factors"; alternatively they are easily obtained from careful
treatment of the double integrals along the same contours).
To restore homogeneity of the prepotential, one needs to
introduce a new modulus, $\Lambda$, but this makes the SW structure
singular, thus more new moduli should be introduced and so on --
often up to the full scaled Whitham theory. As a result, the
set $\{I\}$ of moduli in eq.(\ref{prepgen}) is in fact
infinitely large (includes, at least, all the integrals like
(\ref{Whmoduli})).

Second, eq.(\ref{prepgen}) contains integrals along
the $A$ and $B$ contours, while only $A$-contours occur in 
eq.(\ref{prepdi}). Instead, however, a new object -- the
kernel $G(x,x\prime)$ appears in (\ref{prepdi}).
One can get a clue to understanding  this last phenomenon
by applying the obvious eq.(\ref{prepgen}) to the simplest
example \cite{GMMM}, when the infinite set of additional
moduli is localized at a single point, as described by 
eqs.(\ref{Whmoduli}). Then

\be
\sum_{k=0}^\infty 
\tilde  T_k\frac{\partial{\cal F}}{\partial \tilde T_k} 
= \sum_{k=0}^\infty \frac{1}{k}
\oint_{x=x_0} \oint_{x'=x_0}\frac{(x-x_0)^k}{(x'-x_0)^{k}}
dS(x)dS(x') =
 \nn \\ =  -
\oint_{x=x_0} \oint_{x'=x_0} \log(x-x')dS(x)dS(x')
\label{di2}
\ee 
(the term with $k=0$ is actually logarithmic).
This formula is exactly of the {\it type} (\ref{prepdi}),
with the kernel $G(x,x') = \log(x-x')$.\footnote{
In fact instead of $\log(x-x')$ one could easily get, say
$\log\sinh(x-x')$ or anything else; the formulas (\ref{Whmoduli})
and thus (\ref{di2}) depend on the choice of local coordinates
in the vicinity of $x_0$.
Different types of matrix models (Hermitean, unitary, ...)
with different Van-der-Monde determinants and thus different
kernels $G(x,x')$ are in fact associated with different
possibilities of {\it global} coordinatization of the simple
{\it bare} spectral curves.
}
The next step towards understanding (\ref{prepdi}) should be to
explain why the points $x$ and $x'$ in (\ref{di2}) can belong to
different contours (cuts) when the singular points get distributed
along continuous curves or areas
and when (\ref{prepgen}) and (\ref{di2}) can be generalized to
reproduce (\ref{prepdi}). See the discussion at the end of s.2
above.

\section{Original SW anzatz and 
the Regularized Dijkgraaf-Vafa (DV) setting}

\subsection{The prepotential}

Consider the differential $dS$ of the form

\be
dS = \sqrt{\frac{{\cal P}_{2n}(x)}{{\cal Q}_{2n}(x)}}dx =
\sqrt{\prod_{L=1}^{2n} \frac{x - \beta_L}{x - M_L}}dx
\label{dSreg}
\ee 
Under the usual additional constraint\footnote{
This condition is needed to eliminate the moduli-dependence
of the pole at $x=\infty_{\pm}$ (note that $x=\infty$
describes {\it two} different points at two different sheets
of the hyoerelliptic Riemann surface). 
Instead of imposing (\ref{U1cons})
one can take the polynomial in the numerator to be 
${\cal P}_{2n-1}(x)$ of degree $2n-1$, not $2n$. 
(Note, that exactly such 
$dS_{SU(2)} = \sqrt{\frac{(x-u)}{x(x-M)}}dx$,
with $n=1$ and one of the two $M_L `s$ is equal to zero,
was introduced in the very first example of SW theory in
ref.\cite{SW1}.) With such choice of the numerator the
$2n-1$ moduli $\beta_L$ are unconstrained and some formulas
below are further simplified; e.g. the second term is absent
at the r.h.s. of eq.(\ref{holdef}). However, for the
further discussion of the DV theory we need 
${\cal P}_{2n}(x)$ of degree $2n$.
}

\be
\sum_{L=1}^m \beta_L = 0
\label{U1cons}
\ee 
this expression defines a 
{\it regularized} SW system of the hyperelliptic curves

\be
Y^2 = {\cal P}_{2n}(x){\cal Q}_{2n}(x) =
\prod_{L=1}^{2n} (x - \beta_L)\prod_{L=1}^{2n}(x - M_L)
\label{curva}
\ee 
of genus $2n-1$ with the $2n-1$ ramification points
$\beta_L$ with $L = 1,\ldots,2n-1$ serving as moduli,
and $\beta_{2n}$  expressed through them by (\ref{U1cons}). 
The remaining $2n$ 
ramification points $M_L$, $L = 1,\ldots,2n$ are kept fixed as
external parameters of the family.

A convenient way to choose the $A$- and $B$-cycles 
and the associated {\it flat} moduli is:\footnote{
Actually, it is the one  which is well adjusted for taking the limit of small
$\rho_i$ when $\beta_{2i-1} = \gamma_i - \rho_i$ and
$\beta_{2i} = \gamma_i + \rho_i$. See discussion in s.5 below.
}

\be
S_i = \oint_{A_i} dS = 2\int_{\beta_{2i-1}}^{\beta_{2i}} dS,
\ \ i=1,\ldots, n, 
\nn \\
T_i = \oint_{A_{n+i}} dS = 2\int_{M_{2i-1}}^{M_{2i}} dS,
\ \ i=1,\ldots, n-1
\label{STmodules}
\ee 
and

\be
\frac{\partial {\cal F}(S,T)}{\partial S_i} =
\oint_{B_i} dS = 2\int_{\beta_{2i}}^{M_{2n}} dS,
\ \ i=1,\ldots, n, 
\nn \\
\frac{\partial {\cal F}(S,T)}{\partial T_i} =
\oint_{B_{n+i}} dS = 2\int_{M_{2i}}^{M_{2n}} dS,
\ \ i=1,\ldots, n-1
\ee

The $2n-1$ differentials

\be
dv_L = \frac{\partial dS}{\partial \beta_L} =
-\frac{1}{2}
\left(
\frac{dS}{x - \beta_L} - \frac{dS}{x - \beta_{2n}}
\right), 
\ \ L=1,\ldots, 2n-1
\label{holdef}
\ee 
are all holomorphic, their linear combinations 
with the coefficients from inverted matrix of $A$-periods
provide {\it canonical} 1-differentials $d\omega_L$, 
and the period matrix, made out of
the $B$-periods of $d\omega_L$, is symmetric:

\be
T_{IJ} = \oint_{B_I} d\omega_J = T_{JI}, \ \ \
\oint_{A_I} d\omega_J = \delta_{IJ}, \nn \\
dv_K =  \left(\oint_{A_I} dv_K\right) d\omega_I, \nn \\
\oint_{A_I} dv_K = T_{IJ} \oint_{A_J} dv_K
\label{dvvsdomega}
\ee

\subsection{Residue formula}

Since moduli $\beta_L$ are just the branching points,
the derivation of residue formula in the present case is especially simple.
As usual in the case of hyperelliptic systems, it
is based on the expression
for the {\it hyperelliptic} period matrix derivative over the
position of a branching point \cite{Fay}:

\be
\frac{\partial T_{IJ}}{\partial \beta_L} = 
\hat {\omega}_I(\beta_L) \hat {\omega}_J(\beta_L) -
\hat {\omega}_I(\beta_{2n}) \hat {\omega}_J(\beta_{2n}),
\ \ L=1,\ldots,2n-1.
\label{Tderbp}
\ee 
Here $\hat {\omega}_I(\beta_L)$ denote the {\it value} of
{\it canonical} 1-differential at the branching point $\beta_L$,
i.e. in the vicinity of this point, where $x = \beta_L + \xi_L^2$,
the differential becomes $d\omega(x) = \hat {\omega}_I(\beta_L)d\xi_L
+ O(\xi_L)d\xi_L$. 
The second term at the r.h.s. of eq.(\ref{Tderbp}), 
arises because the constraint (\ref{U1cons}) requires
$\beta_{2n}$ to vary whenever any other $\beta_L$ is changed.

Eq.(\ref{Tderbp}) can be used to fine the third derivative
of the prepotential with respect to the {\it flat} moduli
$\{ a_L \} = \{ S_i, T_i\}$.\footnote{
For the sake of completeness,
we use the chance to correct an impurity in the derivation of residue
formula for the pure $N=2$ SUSY gauge theory in $4d$, which is
present in the (incorrect) formula (64) of Appendix C in 
ref.\cite{MMM1}. Simultaneously we generalize that
proof to include the fundamental matter hypermultiplets.
The relevant family of hyperelliptic curves is given by

$$
y^2 = P_{N_c}^2(x) - Q_{N_f}(x) = 
\prod_{\gamma = 1}^{2N_c}(x-\lambda_\gamma)
$$
Differentiating both sides of this equation, first, by $h_l$
(the coefficients of $P_{N_c}(x)$ and, second, by $x$, and putting
further $x = \lambda_\alpha$, we get respectively:

$$
2P(\lambda_\alpha)\lambda_\alpha^{l-1} = 
-\prod_{\gamma\neq \alpha}(\lambda_\alpha -  \lambda_\gamma)
\frac{\partial \lambda_\alpha}{\partial h_l}
$$ 
and

$$
2P(\lambda_\alpha)P'(\lambda_\alpha) - Q'(\lambda_\alpha) =
\prod_{\gamma\neq \alpha}(\lambda_\alpha -  \lambda_\gamma)
$$
Since at $x=\lambda_\alpha$ we have $y(\lambda_\alpha)=0$ and
$P^2(\lambda_\alpha) = Q'(\lambda_\alpha)$, one can deduce that

$$
\frac{\partial \lambda_\alpha}{\partial h_l} 
= -\frac{\lambda_\alpha^{l-1}}
{2P'(\lambda_\alpha) - \frac{Q'(\lambda_\alpha)}
{P(\lambda_\alpha) + y(\lambda_\alpha)}}
$$
Starting from eq.(65) the derivation of ref.\cite{MMM1}
is already correct, and one arrives at residue formula of the
form

$$
\frac{\partial^3 {\cal F}}
{\partial a_i \partial a_j \partial a_k} =
\sum_{\lambda_{\alpha}}
\frac{\hat {\omega}_i(\lambda_{\alpha})
\hat {\omega}_j(\lambda_{\alpha}) \hat {\omega}_k(\lambda_{\alpha})}
{(P'(\lambda_\alpha) - \frac{1}{2}\frac{Q'(\lambda_\alpha)}
{P(\lambda_\alpha) + y(\lambda_\alpha)})/\hat y(\lambda_\alpha) } 
=
\sum_{dx = 0} res\ 
\frac{d\omega_i d\omega_j d\omega_k}
{ dx d\log (P + y)}
$$
In this case the quadratic differential 
$ddS = dx d\log (P + y) \cong d\lambda d\log w$
is associated with the SW symplectic structure
$d\lambda \wedge d\log w$.
}
The constraint (\ref{U1cons})
makes intermediate formulas look heavier than they actually are, but 
it does not affect explicitly the final expression (\ref{3derFreg}).

\be
F_{IJK} = \frac{\partial^3 {\cal F}}
{\partial a_I \partial a_J \partial a_K} =
\frac{\partial T_{IJ}}{ \partial a_K} =
\sum_{L=1}^{2n-1} \frac{\partial \beta_L}{ \partial a_K} 
\frac{\partial T_{IJ}}{ \partial \beta_L} 
= \nn \\ =
\sum_{L=1}^{2n-1} \frac{\partial \beta_L}{ \partial a_K}
(\hat {\omega}_I(\beta_L) \hat {\omega}_J(\beta_L) -
\hat {\omega}_I(\beta_{2n}) \hat {\omega}_J(\beta_{2n}))
\ee
Actually, expression in brackets at the r.h.s. is equal to

\be
-2\sum_{dS = 0}  res\
\frac{d\omega_I(x) d\omega_J(x) dv_L(x)}
{ dS d\log {\cal P}_{2n}}
\label{dvoverdS}
\ee
where sum is over all zeroes of $dS$ (i.e. over all the points
$\beta_L$, $L = 1,\ldots 2n$). Indeed, the ratio

\be
-2\frac{dv_L(x)}{ dS(x)} = \frac{1}{x-\beta_L} - 
\frac{1}{x-\beta_{2n}}
\ee
has double poles at $\beta_L$ and $\beta_{2n}$ and no other
singulatities, while

\be
d\log {\cal P}_{2n} = \sum_{K=1}^{2n}\frac{dx}{x-\beta_K} 
\ee 
has {\it simple} poles at {\it all} the branching points $\beta_L$
(where $dx$ has simple zeroes). Then the entire ratio in (\ref{dvoverdS})
possesses just two {\it simple} poles at $\beta_L$ and $\beta_{2n}$,
and only these two contribute to the sum with coefficients 
$\hat {\omega}_I(\beta_{L}) \hat {\omega}_J(\beta_{L})$
and $-\hat {\omega}_I(\beta_{2n}) \hat {\omega}_J(\beta_{2n})$
respectively.

It remains to substitute 

\be
\frac{ \partial a_K}{\partial \beta_L} = \oint_{A_K}
\frac{dS}{\partial \beta_L} = \oint_{A_K} dv_L
\ee
and use the relation (\ref{dvvsdomega})
between $dv_L(x)$ and $d\omega_K(x)$, in order to obtain finally the
residue formula for the system (\ref{dSreg}) in the form:

\be
F_{IJK} = \frac{\partial^3 {\cal F}}
{\partial a_I \partial a_J \partial a_K} =
\sum_{dS = 0} res\
\frac{d\omega_I d\omega_J d\omega_K}
{ dS d\log {\cal P}_{2n}} =
- \sum_{d{\cal P}_{2n} = 0} res\
\frac{d\omega_I d\omega_J d\omega_K}
{ dS d\log {\cal P}_{2n}}
\label{3derFreg}
\ee 
In particular,
we see that in the present case the quadratic differential
$ddS = dSd\log {\cal P}_{2n}$, and it is associated in the usual
way with canonical SW symplectic structure

\be
\sum_I d\omega_I(x)\wedge \delta a_I =
\sum_I dv_I(x)\wedge \delta\beta_I =
dS(x) \wedge \delta\log {\cal P}_{2n}(x)
\label{canstr}
\ee

Eq.(\ref{3derFreg}) can be further used for straightforward
derivation of the WDVV equations on the lines of 
\cite{MMM1}-\cite{MMM3}.

\subsection{WDVV equations (derivation)}

Residue formula (\ref{3derFreg}) expresses the prepotential's
third derivatives over {\it flat} moduli, ${\cal F}_{IJK}$,
as an action of certain linear operator, ${\cal L}$, on the
product  $d\omega_I d\omega_J d\omega_K$
of three canonical holomorphic differentials.
Imagine now, that an "algebra" of such differentials can
be defined, i.e. that 

\be
d\omega_I d\omega_J \stackrel{?}{=} C_{IJ}^Ld\omega_L
\label{prod}
\ee
with some coefficients $C_{IJ}^L$. Then, using the linearity of
operation ${\cal L}$,

\be
{\cal F}_{IJK} = {\cal L} \left(d\omega_I d\omega_J d\omega_K\right)
\stackrel{?}{=} C_{IJ}^L {\cal L} \left(d\omega_K d\omega_L\right)
= C_{IJ}^L\eta_{KL}
\ee
or

\be
\left(\check C_I\right)_J^L = C_{IJ}^L =
\check {\cal F}_I \check\eta^{-1}
\ee
If, further, the product in (\ref{prod}) was associative,
i.e. the matrices $\check C_I$ commute,

\be
\check C_I\check C_K = \check C_K\check C_I
\ee
then 

\be
\check {\cal F}_I \check\eta^{-1} \check {\cal F}_K =
\check {\cal F}_K \check\eta^{-1}\check {\cal F}_I 
\label{WDVVderi}
\ee

This is basically the way to derive the WDVV equations.
 It, however, remains to correct two important details.

First, (\ref{prod}) can not be true for 1-forms; instead
one should better write

\be
d\omega_I(x) d\omega_J(x) \stackrel{?}{=} C_{IJ}^L d\omega_L(x)d\eta(x)
\label{prodcor}
\ee
with quadratic differentials at both sides of the equation.
This requires  $d\eta(x)$ to be also a holomorphic 1-differential,

\be
d\eta(x) = \eta_Md\omega_M(x)
\ee
and

\be
\eta_{KL} = {\cal L} \left(d\omega_K d\omega_L d\eta\right)
= \eta_M{\cal F}_{KLM}, \ \
\check\eta = \eta_M \check{\cal F}_M
\ee
If $\eta_M = \delta_{MJ}$ we get from (\ref{WDVVderi}) the
WDVV equations in the form of (\ref{WDVVeqs}).

Second, corrected eq.(\ref{prodcor}) is still too strong
a requirement to be realized in a SW model. Fortunately, we can
add at the r.h.s. any quadratic differential, annihilated by the linear
operation ${\cal L}$. From an explicit form of the residue formula
 (\ref{3derFreg}) it is clear that 

\be
{\cal L}\left(\ldots\cdot d{\cal P}_{2n}\right) = 0
\ee
This means that instead of (\ref{prodcor}) one can demand:

\be
d\omega_I(x) d\omega_J(x) = C_{IJ}^L d\omega_L(x)d\eta(x) +
D_{IJ}^L d\omega_L(x) \frac{d{\cal P}_{2n}(x)}{Y(x)},
\label{prodcorrect}
\ee
while $d\omega_I(x)$ are nothing but polynomials of degree $2n-2$ in $x$
divided by $Y(x)$. Therefore at the l.h.s. of eq.(\ref{prodcorrect})
we can encounter {\it any} polynomial of degree $2(2n-2) = 4n-4$
( on top of $Y^2(x)$ in denominator),
with $4n-3$ different coefficients, while at the r.h.s. we have
exactly $(2n-1) + (2n-2)$ adjustment parameters $C_{IJ}^L$
and $D_{IJ}^L$ to match them. (Note that $D_{IJ}^L d\omega_L(x)Y(x)$
needs to be a polynomial of degree $(4n-4) - (2n-1) = 2n-3$. Thus
there are only $2n-2$ adjustment parameters $D_{IJ}^L$).
This completes the derivation of the WDVV equations in the hyperelliptic
regularized SW model (\ref{dSreg}).\footnote{
Note, that if the $T$-moduli were not taken into account,
and instead one would keep only the differentials which remain
holomorphic after  the transition to the singular DV limit (see 
the next section below), we would actually stay with polynomial of 
degree $n-1$
instead of $2n-2$, as representatives of $d\omega_i$'s, while
residue formula would still contain entire $d{\cal P}_{2n}/dx$
of degree $2n-1$. Then all the coefficients $D_{ij}^l$ would need
to vanish, and just $n$ adjustment parameters $C_{ij}^l$ would not
be sufficient to make the "algebra" closed. Thus there would be
no reason for the WDVV equations to hold, and the singular DV
system, if treated naively, would be {\it incomplete}. It is still
an open question, whether its completeness can be restored
with the help of adequate introduction of $T$-moduli, which can
nicely survive the limiting procedure without destroying generic
properties of SW system.
}

\section{Towards the theory of the CIV-DV prepotential}

\subsection{Lifting regularization}

Original DV setting appears from (\ref{dSreg})
in the specific singular limit, when all 
$M_L \rightarrow \infty$.

In this DV limit the genus $2n-1$ curve 
is degenerated into a pair of genus $n-1$ curves,

\be
y^2(x) = \prod_{L=1}^{2n}(x-\beta_L) 
\label{DVc}
\ee
and

\be
\tilde y^2(x) = \prod_{L=1}^{2n}(x-M_L),
\label{DVinf}
\ee
connected by two long tubes (turning just into two punctures 
at $x=\infty_{\pm}$ on every curve in extreme limit).
The first curve (\ref{DVc})
is where the (unregularized) DV theory lives with 

\be
dS_{DV} = \ {\displaystyle lim_{all \ M_L \rightarrow \infty} } \ dS \cdot
\sqrt{\prod_L (-M_L)}  = y(x) dx,
\label{DVdS}
\ee
while the second curve (\ref{DVinf}) can be used to describe
the regularization procedure in a way, consistent with the SW theory 
(in particular, preserving its {\it completeness}).\footnote{
Note, that the points $x=\infty$ ( the two points on the two
sheets of the hyperelliptic curve), where $M_L$ are prescribed
to condense in the DV limit, are in no way distinguished on
the curve (\ref{curva}); one can shift them to any other position
by a rational transformation of $x$ and $Y(x)$.
However, after such a transformation, the  singular nature
  of  $dS_{DV}$  becomes transparent.
Further introduction of an extra cut-off parameter $\Lambda$
(to make the planar matrix model formally finite),
 does not save the whole situation: this prescription
(its real meaning is to {\it imitate} the splitting between different
$M_L$'s), if formally applied, breaks a lot of the SW structure.
The minimal price, payed by conventional DV prescription,
is to allow canonical differentials with simple poles at $x=\infty$,
and to allow some $\Lambda$-dependent terms (and sometime
even those containing $\sqrt{S}$ without any $\Lambda$-dependence.)
in the prepotential, which  are not taken in subsequent applications.
}

The crucial problem with the limiting procedure is to describe
it in such a way, that the set of $2n-1$ moduli $\beta_L$
of the curve (\ref{DVc}) gets {\it naturally} split into
two parts: the $n$ moduli associated with the 1-differentials
which remain holomorphic on (\ref{DVc})
(in fact, DV recipe requires to
include also a meromorphic differential with {\it simple} poles
at $\infty_\pm$) and the remaining $n-1$ moduli -- the remnants
of our $T_i$ which are not considered in the conventional DV approach.
Such splitting is not automatically guaranteed by our
separation (\ref{STmodules}) of $S$ and $T$-moduli in 
{\it regularized} theory,  despite that the limit of $S$'s
is definitely correct.\footnote{
To give just one example of non-trivial alternative, define
$S$'s and $T$'s in (\ref{STmodules}) with the help of another
basis of canonical cycles: take instead of
$A_i$ (from $\beta_{2i-1}$ to $\beta_{2i}$),
$\tilde A_i$ (from $M_{2i-1}$ to $M_{2i}$),
$B_i$ (from $\beta_{2i}$ to $M_{2n}$),
$\tilde B_i$ (from $M_{2i}$ to $M_{2n}$)
the symplectically conjugate set, consisting of
$A_i$, $\tilde A_i - A_i$, $B_i+\tilde B_i$, $\tilde B_i$.
Then the definition of the $S$ moduli remains unchanged, but
those of $T$'s and, notably, of $\partial {\cal F}/\partial S_i$
(now defined with {\it new} $T$'s kept fixed under differentiation)
become different. The definitions can still coincide in the limit,
in particular,
the difference in the formulas for $\partial {\cal F}/\partial S_i$
can seem not too big: 
these are given by integrals from $\beta_{2i}$ to one
and the same point $M_{2n}$ if one uses the basis (\ref{DVc}),
and from $\beta_{2i}$
to the $i$-dependent point $M_{2i}$ if the alternative basis is
used. In the limit when all $M_i \rightarrow \infty$ these 
prescriptions do not seem to differ too much, but since the limit
is highly singular, they actually can lead to very different formulas.
}
The next question is whether any information about the $T$-moduli
survives  the limit, ( which should be because these moduli before the limit
certainly survive as {\it parameters} in the DV setting), and
whether the formulas for prepotential's $T$-derivatives and 
residue formula for its third derivatives can also be preserved.

Full solution of this problem remains for the future research
(which can actually reveal different universality classes
with different prepotentials, corresponding to different degenerations
of the  "hidden-sector" curve (\ref{DVinf}) in the limit).
The rest of this section contains comments on variety of
notation, useful for attacking this kind of problem,
and on its simplest thinkable solution. Potential problem 
with such solution (and instead confirming the option of
different universality classes) will be illustrated in a separate
publication \cite{IM5}.

\subsection{Various parametrizations of the DV family}

On the "hidden-sector" curve (\ref{DVinf}), $dS$ can be represented as

\be
dS = \frac{dx}{\tilde y(x)}\sum_{k=0}^\infty u_{k-n} x^{n-k}
\ee
The sum here comes from the large-$x$ expansion of 
$\sqrt{{\cal P}_{2n}}(x)$:

\be
\sqrt{{\cal P}_{2n}(x)} = \sum_{k=0}^\infty u_{k-n} x^{n-k}
= P_n(x) + \sum_{k=1}^\infty u_k x^{-k}
\ee
The new polynomial 

\be
P_n(x) = \prod_{i=1}^n (x-\alpha_i),
\label{Pn}
\ee
which naturally emerges here, is of
degree $n$.\footnote{
It is this polynomial that  is identified
within the DV theory \cite{DV} with the derivative of
the polynomial {\it superpotential} $W_{n+1}(x)$
of the $N=1$ SUSY theory in $4d$, which, in turn, provides 
an action of $d=0$ matrix model, describing the average over
constant fields:

$$
P_n(x) = \frac{1}{g_{n+1}}W'_{n+1}(x) =
\prod_{i=1}^n (x - \alpha_i), \ \ 
W_{n+1}(x) = g_{n+1}\left(x^{n+1} + O(x^{n-1})\right)
$$
}
Its roots $\alpha_i$ are functions of $\beta_L$,
subjected to the usual $U(1)$ constraint

\be
\sum_{i=1}^n \alpha_i = 0,
\label{U1dec}
\ee
and the remaining $(2n-1) - (n-1) = n$ moduli can be encoded
-- in this new parametrization -- either by the first $n$
of the variables $u_k$ (i.e. by $u_1,u_2 \ldots, u_n$) 
or by the $n$ coefficients of still another polynomial $f_{n-1}(x)$,
of degree $n-1$, accounting for the difference between 
${\cal P}_{2n}(x)$ and $P_n^2(x)$:

\be
y^2(x) = {\cal P}_{2n}(x) = P_n^2(x) + f_{n-1}(x) =
P_n^2(x) + u_1Q_{n-1}(x)
\label{curvedef}
\ee
Here 

\be
Q_{n-1}(x) = x^{n-1} + O(x^{n-2}) =
\prod_{\nu = 1}^{n-1}(x - m_\nu)
\ee 
has unit coefficient in front of the $x^{n-1}$  term 
(while the  coefficient in the corresponding term  in $f_{n-1}(x)$ is $u_1$). 
Also, its root decomposition reminds  us of another
example of SW theory \cite{Masf}, the one 
gauge theories with massive hypermultiplets
in the fundamental representation of the gauge group
(in this particular case $N_c = n$ and $N_f = N_c-1$).

This parallel implies still 
another parametrization of the same DV curve (\ref{curvedef}):

\be
w + \frac{Q_{n-1}(x)}{w} = 2P_n(x), \nn \\
w - \frac{Q_{n-1}(x)}{w} = 2y(x)
\label{wlpar}
\ee
The difference between the two versions of SW theory is that
for the $N=2$ SUSY gauge theory in $4d$
the SW differential $dS_4 \cong \log w dx$ 
(for its $5d$ counterpart \cite{5d}, instead,
$dS_5 \cong \log w d\log x$), while

\be
dS_{DV} = y(x)dx \cong wdx
\label{dSDVdef}
\ee
Discussion of the obviously emerging chain of differentials
 $dS_{DV}$, $dS_4$, $dS_5$
remains beyond the scope of the present paper.

Yet another representation of the same polynomial
$f_{n-1}(x) = u_1Q_{n-1}(x)$,
convenient for the study of {\it perturbative}
(logarithmic) contribution to the prepotential, is

\be
f_{n-1}(x) = P_n(x)\sum_{i=1}^n\frac{\tilde S_i}{x-\alpha_i}
= \sum_{i=1}^n \tilde S_i\prod_{j\neq i}^n (x-\alpha_j)
\ee
In perturbative limit $\tilde S_i = S_i + O(S^2)$ and

\be
dS^{pert}_{DV} = P_n(x)dx + \sum_{i=1}^n\frac{S_i}{x-\alpha_i} dx.
\ee

Significance of parameters $\alpha_i$ is in that {\it these} are
the ones which are kept constant in the definition of conventional
(unregularized) CIV-DV \cite{CIV,DV} prepotential,
${\cal F}(S_i|\alpha_i)$, 

\be
S_i = \oint_{A_i} dS_{DV}, 
\nn \\
\left.\frac{\partial{\cal F}}
{\partial S_i}\right|_{constant\ \alpha's} = 
\int_{\gamma_i + \rho_i}^\Lambda dS_{DV}, 
\label{CIVDVprep}
\ee
$i = 1,\ldots,n$.
The main problem of taking the
DV limit from the regularized case is to identify the $T$-moduli
in such a way (ways?)
that condition of $T$'s being fixed smoothly transforms in the limit
into condition of $\alpha$'s being fixed).

Along with the points $\alpha_i$ it is useful to introduce another set
of points, namely, the positions $\gamma_i$ of the cut centers, so that

\be
y^2 = \prod_{L=1}^{2n}(x-\beta_L) =
\prod_{i=1}^n ((x-\gamma_i)^2 - \rho_i^2)
\ee
i.e. ramification points $\{\beta_L\} = \{\gamma_i \pm \rho_i\}$,
and 

\be
\sum_{i=1}^n \gamma_i = 0.
\ee
 An advantage of this parametrization is that for
 the small cut lengths $\rho_i$
the {\it flat} moduli $S_i$ are also small, and expansion of the 
prepotential in powers of $S_i$ is related to its expansion in powers
of $\rho_i$ (actually, $\rho_i^2$).  This later one is rather
 straightforward to
obtain, and can serve as the first step towards the calculation of
${\cal F}(S|\alpha)$.

\subsection{Calculating the prepotential}

Evaluation of the (unregularized) CIV-DV prepotential 
is a tedious procedure, involving the following steps.

1. First of all, one needs to evaluate the integral 

\be
\int dS_{DV}(x) =
\int \sqrt{{\cal P}_{2n}(x)} dx =
\int
\prod_{j=1}^n \sqrt{(x-\gamma_j)^2 - \rho_j^2}\ dx
\ee
as a function of $\gamma$'s and $\rho$'s.
The answer can be found in the form of a power series in
$\rho_j^2$ with sophisticated $\gamma$-dependent coefficients,
calling for a representation theory interpretation.
Definite integrals between $\gamma_i-\rho_i$ and
$\gamma_i+\rho_i$ provide $S_i/2g_{n+1}$,  while integrals between
$\gamma_i+\rho_i$ and $\Lambda$ 
(where $\Lambda$ is somewhat like the common gathering point for
all $M_L$'s on their way to infinity)
provide $\frac{1}{2g_{n+1}}\left.\frac{\partial{\cal F}}
{\partial S_i}\right|_{constant\ \alpha's}$.
In order to integrate these formulas and obtain 
${\cal F}(S|\alpha)$, one now needs to switch from
$\gamma$'s to $\alpha$'s.

2. The expression of $\gamma$'s  by $\alpha$'s and $\rho$'s
arises from  comparison of the coefficients in front of
$x^{2n-2}, x^{2n-3},\ldots, x^n$ at both sides of the equation

\be
\prod_{i}^n \left((x-\gamma_j)^2 - \rho_j^2\right) =
\left[\prod_{i}^n (x-\alpha_j)\right]^2 + f_{n-1}(x)
\ee
It is easy to see that $\sigma_i \equiv \gamma_i - \alpha_i
= O(\rho^2) = O(S)$.
After solving this system of $n-1$ equations (in the form of
infinite series in $\rho$'s, sometime summable), 
expressions for $\gamma$'s can be substituted into the
formulas for $S_i$ and ${\partial{\cal F}}/{\partial S_i}$,
derived at the previous step.

3. Now expressions for $S_i(\rho^2|\alpha)$ should be inverted
to provide $\rho_i^2(S | \alpha)$, which are further substituted
into the formula for ${\partial{\cal F}}/{\partial S_i}$
to provide these
derivatives in the form of expansion in powers of $S$'s with
coefficients, made out of $\alpha$'s (which presumably still
have a straightforward representation-theory interpretation).\footnote{
In fact, it appears to be that step 1 and step 2 have closer relation to
representation theory than the prepotential itself (relation can
be spoiled at step 3). It can happen that  the connection of the
 prepotential theory to the
finite-$N$ matrix models and ordinary $\tau$-functions 
 can be found more easily at the level of $S(\gamma,\rho)$ or 
$S(\alpha,\rho)$ (not a big surprise, since the prepotential itself,
being a {\it quasiclassical} $\tau$-function, does not need to have
a transparent group-theoretical structure). A similar (related?)
phenomenon can be observed in \cite{Ne}.
}

The resulting expressions are already easy to integrate and we
obtain ${\cal F}(S|\alpha)$ in a form of expansion in
powers of $S$'s with coefficients made out of $\alpha$'s:

\be
2\pi i{\cal F}(S|\alpha) = 4\pi ig_{n+1}\left(W_{n+1}(\Lambda)\sum_i S_i 
- \sum_i W_{n+1}(\alpha_i)S_i\right) -
(\sum_{i} S_i)^2\log\Lambda +
\nn \\ +
\frac{1}{2} \sum_{i=1}^n S_i^2\left(\log \frac{S_i}{4}\ - \frac{3}{2}\right) 
- \frac{1}{2}\sum_{i<j}^n(S_i^2 - 4S_iS_j + S_j^2)\log \alpha_{ij} + 
\sum_{k=1}^\infty \frac{1}{(i\pi g_{n+1})^k}{\cal F}_{k+2}(S|\alpha),
\label{prepoexp}
\ee
where ${\cal F}_{k+2}(S|\alpha)$ are polynomials of degree $k+2$ in $S$'s
with $\alpha$-dependent coefficients,
and $W_{n+1}'(x) = P_n(x)$.

\subsection{$T$-moduli}

We suggest to supplement these 3 steps by the final one
aimed at replacing the $\alpha$ dependence by that
on the {\it flat} moduli $T_i$, $i=1,\ldots,n-1$.
An immediate problem, as we already discussed, is to find 
an adequate definition of these in the singular DV limit.

A possible guess could be to take just

\be
\tilde T_k = res_\infty  x^{-k} dS_{DV}
= res_\infty  x^{-k} P_n(x)dx 
= \nn \\ = e_{n-k-1}(\alpha) \equiv
(-)^{n-k-1} \sum_{i_1<\ldots<i_{n-k-1}}\alpha_{i_1}\ldots
\alpha_{i_{n-k-1}},
\label{Tdef}
\ee
$k = 1,\ldots,n-1$ (obviously,  $T_0 = \sum_{i=1}^n S_i$ would be the case) and

\be
\frac{\partial{\cal F}}{\partial \tilde T_k} = 
\frac{1}{k}res_\infty  x^{+k} dS_{DV}.
\label{Tdef2}
\ee
A problem with such an ansatz is that individual
differences $\alpha_{ij} = \alpha_i-\alpha_j$, appearing
in (\ref{prepoexp}), are difficult to express through 
symmetric polynomia of $\alpha$'s such as $e_{n-k-1}(\alpha)$.
A somewhat more sophisticated guess
based on the previous work with Whitham
hierarchies \cite{GMMM} can be to replace $x^{\pm k}$
in these formulas by $w^{\pm k/n}$ with $w$ defined in
(\ref{wlpar}).

Whatever the right formulas are, they should obey consistency
conditions (
$\partial^2{\cal F}/\partial S_i\partial \tilde T_j$ and
$\partial^2{\cal F}/\partial \tilde T_i\partial \tilde T_j$
  being symmetric)
and the proper form of the residue formula for the third
derivatives should be found. Afterwards one can proceed to the
WDVV equations, which can hypothetically survive after
the DV limit is taken.

\section{Conclusion}

To conclude, we outlined a program for the study of DV
prepotential from the perspective of SW theory.
Its main non-trivial ingredients include the need to
 study the {\it regularized} DV theory (\ref{dSreg});
introduce additional $T$-moduli and relate them
to the ones arising from simple \cite{IM2,GMMM} and
sophisticated \cite{KMWZ} Whitham hierarchies; 
to check the validity of the WDVV equations (and, probably,
discover  their adequate generalization to elliptic
situation); and to find a representation-theory
interpretation of periods (and may be even the prepotential). 
If fulfilled, this
program can provide non-trivial conceptual clues 
for developments of theory of effective-action as well as
 helpful machinery to some more concrete questions such as relations to
finite-$N$ matrix models \cite{UFN3}, KP/Toda 
$\tau$-functions \cite{tauf},  and
instanton calculus of \cite{LMNS}-\cite{F}.

\section{Acknowledgements}

A.M. acknowledges the support of JSPS and the hospitality
of the Osaka City University during his stay at Japan.
Our work is partly supported by the 
Grant-in-Aid for Scientific Research (14540284) from the
Ministry of Education, Science and Culture, Japan (H.I),
and by the Russian President's grant 00-15-99296, RFBR-01-02-17488,
INTAS 00-561 and by Volkswagen-Stiftung (A.M.).

\end{document}